\documentclass{elsartL}
\usepackage{graphicx}
\usepackage{amsmath}
\begin{document}

\begin{frontmatter}
\title{New analysis of the low-energy $\pi^\pm p$ differential cross sections of the CHAOS Collaboration}
\author[EM]{E. Matsinos{$^*$}},
\author[GR]{G. Rasche},
\address[EM]{Institute of Mechatronic Systems, Zurich University of Applied Sciences, Technikumstrasse 5, CH-8401 Winterthur, Switzerland}
\address[GR]{Physik-Institut der Universit\"at Z\"urich, Winterthurerstrasse 190, CH-8057 Z\"urich, Switzerland}

\begin{abstract}
In a previous paper, we reported the results of a partial-wave analysis of the pion-nucleon ($\pi N$) differential cross sections (DCSs) of the CHAOS Collaboration and came to the conclusion that the angular distribution of their $\pi^+ p$ 
data sets is incompatible with the rest of the modern (meson-factory) database. The present work, re-addressing this issue, has been instigated by a number of recent improvements in our analysis, namely regarding the inclusion of the 
theoretical uncertainties when investigating the reproduction of experimental data sets on the basis of a given `theoretical' solution, modifications in the parameterisation of the form factors of the proton and of the pion entering the 
electromagnetic part of the $\pi N$ amplitude, and the inclusion of the effects of the variation of the $\sigma$-meson mass when fitting the ETH model of the $\pi N$ interaction to the experimental data. The new analysis of the CHAOS 
DCSs confirms our earlier conclusions and casts doubt on the value for the $\pi N$ $\Sigma$ term, which Stahov, Clement, and Wagner have extracted from these data.
\end{abstract}
{$^*$}{Corresponding author. E-mail: evangelos[DOT]matsinos[AT]zhaw[DOT]ch, evangelos[DOT]matsinos[AT]sunrise[DOT]ch}
\end{frontmatter}

\section{\label{sec:Introduction}Introduction}

Stahov, Clement, and Wagner \cite{scw} have evaluated the pion-nucleon ($\pi N$) $\Sigma$ term from the $\pi^\pm p$ differential cross sections (DCSs) of the CHAOS Collaboration \cite{chaos,denz}; the extracted value was $59 \pm 12$ 
MeV. In Ref.~\cite{mr1}, which was available online (Nucl.~Phys.~A web site) a few months prior to the publication of their paper \cite{scw}, we had reported the details of a partial-wave analysis (PWA) of the same data (DENZ04, in our 
notation). To avoid influence from extraneous sources, an exclusive analysis of the DENZ04 DCSs had been performed in Ref.~\cite{mr1}, after applying to these data the same analysis criteria which had been used earlier \cite{mr2} in a 
PWA of the rest of the low-energy (pion laboratory kinetic energy $T \leq 100$ MeV) $\pi^\pm p$ elastic-scattering measurements.

In the first part of our paper \cite{mr1} (optimisation), we had analysed the DENZ04 DCSs following two theoretical approaches, one featuring standard low-energy parameterisations of the $s$- and $p$-wave $K$-matrix elements, the other 
using the $s$- and $p$-wave $K$-matrix elements of the ETH model~\footnote{The ETH model of the $\pi N$ interaction contains $t$-channel $\sigma$- and $\rho$-exchange graphs, as well as the $s$- and $u$-channel contributions with all 
the well-established $s$ and $p$ baryon states with masses below $2$ GeV; the model amplitudes obey crossing symmetry and isospin invariance.} \cite{glmbg}. Both ways failed to produce acceptable results. We subsequently investigated 
the reproduction of the DENZ04 DCSs on the basis of the results of Ref.~\cite{mr2}. The comparison revealed large discrepancies in the DENZ04 $\pi^+ p$ data sets at forward and medium scattering angles, at all five energies covered by 
the CHAOS experiment. We thus concluded that the angular distribution of the DENZ04 $\pi^+ p$ DCSs was incompatible (in shape) with the rest of the modern (meson-factory) low-energy database.

Owing to the fact that the present paper reports the results of a new analysis of the DENZ04 DCSs, the overlap with the material of Ref.~\cite{mr1} is inevitable. There are three main reasons for embarking on a new analysis of these 
measurements.
\begin{itemize}
\item Included now in our results, for the first time, are also the uncertainties of the theoretical values ($\delta y_{ij}^{th}$) when investigating the reproduction of experimental data sets on the basis of a given `theoretical' 
solution (from now on, `baseline solution' or BLS); these uncertainties are sizeable in some kinematical regions covered by the CHAOS experiment, e.g., in backward $\pi^- p$ elastic scattering. It is worth noting that, compared to our 
past PWAs, the uncertainties $\delta y_{ij}^{th}$ are now somewhat larger, as they also contain the effects of the variation of the $\sigma$-meson mass $m_\sigma$ (see below).
\item Recent developments regarding the proton electromagnetic form factors suggest the replacement of the forms we have used earlier. The parameterisation of the Dirac $F_1^p (t)$ and the Pauli $F_2^p (t)$ 
form factors of the proton with (traditional) dipole forms has been found to provide a poor description of the `world' electron-proton ($e p$) unpolarised and polarised data \cite{bern}. Although the sensitivity of our results to 
the details of the parameterisation of these quantities is low (due to the smallness of the $Q^2$ transfer at low energies), we will nevertheless adopt a more recent parameterisation scheme. In Ref.~\cite{vamx}, the authors made 
use of the so-called Pad\'e parameterisation \cite{amt} for the Sachs electromagnetic form factors $G_E^p (t)$ and $G_M^p (t)$ (in Ref.~\cite{vamx}, the superscript $p$ is omitted), and obtained the optimal values for the relevant 
parameters from a fit to $e p$ measurements (see their Table II). The value of the proton rms electric charge radius $\sqrt{<r_e^2>}$, evaluated from the content of that table (in fact, from the values of the parameters $q_2$ and $q_6$), 
is almost identical to the result of Ref.~\cite{mtn}, thus disagreeing with the $\sqrt{<r_e^2>}$ value extracted from muonic hydrogen \cite{ant}. From now on, we will use the results of Ref.~\cite{vamx} in our analysis software. The pion 
form factor $F^\pi (t)$ is usually parameterised via a monopole form, e.g., see Ref.~\cite{na7}. Despite the fact that, in the low-energy region, results of similar quality are obtained with either a monopole or a dipole form, we will 
adopt the monopole parameterisation henceforth.
\item The fits of the ETH model to the experimental data now involve the variation of $m_\sigma$ within the interval which is currently recommended by the Particle-Data Group (PDG) \cite{pdg}, namely between $400$ and $550$ MeV. The 
earlier fits were made assuming a fixed $m_\sigma$ value ($860$ MeV \cite{torn}).
\end{itemize}
There are three additional reasons, albeit less important, for revisiting this subject. a) The $s$-channel contribution of the Roper resonance to the $K$-matrix element $K_{1-}^{1/2}$, as given in Section 3.5.1 of Ref.~\cite{mr3}, is 
now explicitly included in the first step of the optimisation (low-energy parameterisations of the $s$- and $p$-wave $K$-matrix elements); this change induces very small effects in the fits to the $\pi^- p$ elastic-scattering data. b) 
An improved approach for fixing the (small) $d$ and $f$ waves has been implemented; to suppress artefacts which are due to the truncation of small values, simple polynomials are now fitted to the $d$- and $f$-wave phase shifts of the 
current solution of the SAID analysis \cite{abws}. c) Used in Refs.~\cite{mr1,mr2} were the results of an earlier compilation of the physical constants by the PDG; the results of the most recent compilation \cite{pdg} are used herein.

The present paper is organised as follows. In Section \ref{sec:Method}, we give the details regarding the assessment of the goodness of the reproduction of an experimental data set on the basis of a BLS. The tests are now optimally 
structured and may be generally used in such investigations; for instance, we can use the prescribed tests to assess the reproduction of the $\pi^- p$ charge-exchange ($\pi^- p\rightarrow \pi^0 n$) data on the basis of 
elastic-scattering results (pursuing the investigation of the violation of the isospin invariance in the hadronic part of the $\pi N$ interaction). Section \ref{sec:Results} contains the results we obtained from the new analysis of 
the DENZ04 DCSs. We briefly discuss the implications of our findings in Section \ref{sec:Conclusions}.

\section{\label{sec:Method}Assessment of the goodness of the reproduction of an experimental data set on the basis of a BLS}

Prior to advancing to the technical details, we will give a few definitions which are of relevance in the discussion; if introduced later, they might disrupt the smooth description of the three tests delineated in the present section. 
For our purposes, a BLS is defined as follows.
\begin{itemize}
\item A BLS is a set of values and associated uncertainties ($y_{ij}^{th}$, $\delta y_{ij}^{th}$, $i \in \left[ 1,N_j \right]$), corresponding to the values of the kinematical variables, i.e., of the centre-of-mass scattering angle 
$\theta$ and of $T$, at which the experimental data ($y_{ij}^{exp}$, $\delta y_{ij}^{exp}$, $i \in \left[ 1,N_j \right]$) have been acquired. The indices $i$ and $j$ identify the particular measurement, namely as the $i^{\rm th}$ data 
point of the $j^{\rm th}$ data set. The number of data points of the $j^{\rm th}$ data set is denoted as $N_j$.
\item A BLS comprises predictions obtained via a Monte-Carlo simulation taking into account the results of the optimisation (i.e., the fitted values and the uncertainties of the model parameters, as well as the covariance matrix of each 
fit) of a PWA of $\pi N$ data.
\end{itemize}

Being a sum of independent normalised residuals, each following the normal distribution, our test-statistic is expected to follow the $\chi^2$ distribution. As we will concentrate on this distribution hereafter, aiming at the 
identification of data sets which are poorly reproduced, we will tailor all expressions of the present section to \emph{one-sided} tests (right-tail events).

Let us assume that the background process, underlying the phenomenon under investigation, is a stochastic one (or, equivalently, that the null hypothesis is valid), described by the probability density function $f(x) \geq 0$, where 
$x \in [0,\infty)$ is a numerical result obtained via a measurement made on the observed system. Kolmogorov's second axiom dictates that
\begin{equation} \label{eq:EQ00}
\int_{0}^\infty f(x) dx = 1 \, \, \, .
\end{equation}
The so-called p-value~\footnote{It is casual to refer to p-values in the statistical hypothesis testing in most domains of basic or applied research in economics, psychology, biology, medical physics, etc.} is defined as the upper tail 
of the corresponding cumulative distribution function:
\begin{equation} \label{eq:EQ01}
{\rm p} (x_0) = \int_{x_0}^\infty f(x) dx \, \, \, .
\end{equation}
The p-value is the probability that a measurement, obtained from the observed system, yield a result $x$ which is more statistically significant than $x_0$ (in our case, that $x>x_0$). Assuming the validity of the null hypothesis, the 
p-value may thus be used as a measure of the result $x_0$ being due to chance: `small' p-values attest to the statistical significance of that measurement.

Of course, before assessing the statistical significance of a measurement, one must define what is meant by `small' p-values. Unfortunately, the threshold signifying the outset of statistical significance is subjective; in reality, the 
setting of the significance level~\footnote{The significance level is usually denoted as $\alpha$ in Statistics.} $\mathrm{p}_{min}$ rests on a delicate trade-off between two risks: a) of accepting the alternative hypothesis (of an 
effect not being due to statistical contrivance) when it is false and b) of rejecting the alternative hypothesis when it is true. Of relevance in the choice of the $\mathrm{p}_{min}$ value is which of these two risks is being assigned 
greater importance. For instance, if the implications of risk (b) are considered to be more severe, compared to those of risk (a), an increase of the $\mathrm{p}_{min}$ value is tenable.

Most statisticians accept $\mathrm{p}_{min}=10^{-2}$ as the outset of statistical significance and $\mathrm{p}_{min}=5 \cdot 10^{-2}$ as the threshold indicating probable significance. An interesting recent paper \cite{jhn} interprets 
the lack of reproducibility of scientific results in various disciplines as evidence that the currently-accepted $\mathrm{p}_{min}$ values are rather `optimistic'; the author thus recommends the reduction of these thresholds by one order 
of magnitude~\footnote{Although Ref.~\cite{jhn} states that ``nonreproducibility in scientific studies can be attributed to a number of factors, including poor research designs, flawed statistical analyses, and scientific misconduct'', 
we believe that, at least as far as the research in $\pi N$ physics is concerned, the main reason might simply be `excessive optimism' when assessing the systematic effects in the experiments; in all probability, these uncertainties are 
frequently, if not systematically, underestimated.}.

The probability density function of the $\chi^2$ distribution with $\nu>0$ degrees of freedom (DOF) is
\begin{equation} \label{eq:EQ02}
f(x,\nu) = \begin{cases} \frac{1}{2^{\nu/2} \Gamma(\nu/2)} x^{\nu/2-1} \exp(-x/2), & \mbox{for } x \geq 0 \\ 0 , & \mbox{otherwise} \end{cases} \, \, \, ,
\end{equation}
where $\Gamma(y)$ is the standard gamma function:
\begin{equation} \label{eq:EQ03}
\Gamma (y) = \int_{0}^\infty t^{y-1} \exp(-t) dt \, \, \, .
\end{equation}
For a quantity $x$ following the $\chi^2$ distribution, the expectation value $E[x]$ is simply equal to $\nu$ and the variance $E[x^2]-(E[x])^2$ is equal to $2 \nu$. The relation $E[x]=\nu$ has led many physicists to the use of the reduced 
$\chi^2$ value (i.e., of the ratio $\chi^2/\nu$) when assessing the goodness of the data description in modelling; as long as $\chi^2/\nu \approx 1$, the results are claimed to be satisfactory. Of course, the interesting question in the 
statistical hypothesis testing relates to the value of $\chi^2/\nu$ at which the results start appearing \emph{un}satisfactory; evidently, a threshold value for $\chi^2/\nu$ may be extracted from $\mathrm{p}_{min}$, yet it turns out to 
be $\nu$-dependent, hence cumbersome to use. Such a departure from simplicity is meaningless. It makes more sense to perform the direct test and assess the statistical significance by comparing the p-value, associated with the observed 
$\chi^2$ for $\nu$ DOF, with $\mathrm{p}_{min}$; this is achieved by simply inserting $f(x,\nu)$ of Eq.~(\ref{eq:EQ02}) into Eq.~(\ref{eq:EQ01}), along with $x_0=\chi^2$, and evaluating the integral; several software implementations of 
dedicated algorithms are available, e.g., see Refs.~\cite{as} (Chapter on `Gamma Function and Related Functions') and \cite{ptvf}, the routine PROB of the CERN software library, the functions CHIDIST/CHISQ.DIST.RT of Microsoft Excel, etc.

One additional `side' remark is due. Various definitions of the `data set' have been in use, involving different choices of the experimental conditions which must remain stable/constant during the data-acquisition session. The properties 
of the incident beam, as well as the (physical, geometrical) properties of the target, have been used in the past in order to distinguish the data sets of experiments performed at one place (i.e., at a meson factory) over a short period 
of time (typically, a few weeks). However, data sets have appeared in experimental reports relevant to the $\pi N$ system, which not only involved different beam energies, but also contained measurements of different reactions (e.g., 
mixing $\pi^+ p$ \emph{and} $\pi^- p$ elastic-scattering measurements). As a result, the only prerequisite for accepting measurements as comprising one data set is that they share the same absolute normalisation (and, consequently, 
normalisation uncertainty $\delta z_j$). Of course, this is only a prerequisite, hence a necessary, not a sufficient, condition. The decision regarding the acceptance of a set of measurements as comprising one data set cannot be made 
without an investigation of the stability of the experimental conditions at which the raw measurements had been acquired (this may be difficult to assess), as well as of their (on-line and off-line) processing on the way to the extraction of 
the final experimental results.

We now enter the details of the reproduction. Let us assume that the absolute normalisation of the $j^{\rm th}$ data set is known up to a relative uncertainty $\delta z_j$. Let us also assume that none of the important quantities, 
appearing in the denominators of the expressions of the present section, vanishes. (For the sake of compatibility with our past works \cite{mr1,mr2,mworg}, we will retain the index $j$ in the expressions, despite the fact that its use 
in this section is redundant.)

One way of assessing the goodness of the reproduction of the $j^{\rm th}$ data set involves the determination of the amount of scaling (application of a multiplicative factor to the BLS, enabling the increase or decrease of its 
values, resulting in its `upward' or `downward' shift as `one piece') which must be applied to the BLS ($y_{ij}^{th}$, $\delta y_{ij}^{th}$, $i \in \left[ 1,N_j \right]$) in order that it `best' accounts for the entire data set 
($y_{ij}^{exp}$, $\delta y_{ij}^{exp}$, $i \in \left[ 1,N_j \right]$). Regarding the reproduction of data sets by a BLS, we have gained inspiration from the Arndt-Roper formula \cite{ar}, which we have been using in our fits to the 
experimental data since Ref.~\cite{mworg}. We will next propose tests of the overall reproduction, of the shape, and of the absolute normalisation of the data sets.

We first evaluate the ratios, $r_{ij}=y_{ij}^{exp}/y_{ij}^{th}$; if the quantities $y_{ij}^{exp}$ and $y_{ij}^{th}$ are independent (which is certainly true in our case, as the DENZ04 measurements have not been used in the determination 
of the BLS), the uncertainties $\delta r_{ij}$ are obtained via the application of Gauss's error-propagation formula:
\begin{equation} \label{eq:EQ04}
\delta r_{ij}=r_{ij} \sqrt{\left( \frac{\delta y_{ij}^{exp}}{y_{ij}^{exp}} \right)^2 + \left( \frac{\delta y_{ij}^{th}}{y_{ij}^{th}} \right)^2} \, \, \, .
\end{equation}

The goodness of the reproduction is assessed on the basis of the function $\chi^2_j (z_j)$ defined as:
\begin{equation} \label{eq:EQ05}
\chi^2_j (z_j) = \sum_{i=1}^{N_j} \left( \frac{r_{ij} - z_j}{\delta r_{ij}} \right)^2 + \left( \frac{z_j - 1}{\delta z_j} \right)^2 \, \, \, .
\end{equation}
It is convenient to introduce the weights $w_{ij}$ via the relation $w_{ij} = (\delta r_{ij})^{-2}$.

The second term on the right-hand side of Eq.~(\ref{eq:EQ05}) takes account of the application of scaling to the BLS. This contribution depends on how well the absolute normalisation of the $j^{\rm th}$ data set is known. For a 
poorly-known absolute normalisation, $\delta z_j$ is large and the resulting contribution from the scaling is small; the opposite is true in case of a well-known absolute normalisation. Evidently, the `best' reproduction of the 
$j^{\rm th}$ data set is achieved when, by varying the scale factor $z_j$, the function $\chi^2_j (z_j)$ is minimised:
\begin{equation} \label{eq:EQ06}
\frac{\partial \chi^2_j (z_j)}{\partial z_j} = 0 \, \, \, .
\end{equation}
The solution of this equation is
\begin{equation} \label{eq:EQ07}
z_j = \frac{\sum_{i=1}^{N_j} w_{ij} r_{ij} + (\delta z_j)^{-2}}{\sum_{i=1}^{N_j} w_{ij} + (\delta z_j)^{-2}} \, \, \, .
\end{equation}
Inserting this expression for $z_j$ into Eq.~(\ref{eq:EQ05}), one obtains
\begin{align} \label{eq:EQ08}
(\chi^2_j)_{min} = & \frac{1}{\sum_{i=1}^{N_j} w_{ij} + (\delta z_j)^{-2}} \Big( \sum_{i=1}^{N_j} w_{ij} \sum_{i=1}^{N_j} w_{ij} r_{ij}^2 - \big( \sum_{i=1}^{N_j} w_{ij} r_{ij} \big)^2 \nonumber \\
& + (\delta z_j)^{-2} \sum_{i=1}^{N_j} w_{ij} (r_{ij} - 1)^2 \Big) \, \, \, .
\end{align}

Expression (\ref{eq:EQ08}) yields the minimal $\chi^2$ value in the description of the $j^{\rm th}$ data set, containing $N_j$ data points. In fact, one additional measurement had been made on this data set, namely the one fixing its 
absolute normalisation, which is known with relative uncertainty $\delta z_j$. As a result, the number of DOF for this data set is equal to $N_j + 1 - 1 = N_j$; the subtraction of one unit is due to the use of Eq.~(\ref{eq:EQ07}) as a 
constraint, fixing the value of the scale factor $z_j$. Therefore, the quantity $(\chi^2_j)_{min}$ of Eq.~(\ref{eq:EQ08}) is expected to follow the $\chi^2$ distribution with $\nu = N_j$ DOF. To obtain the p-value of the overall 
reproduction of the $j^{\rm th}$ data set, one uses Eq.~(\ref{eq:EQ01}) with $f(x)=f(x,\nu)$ of Eq.~(\ref{eq:EQ02}), along with $x_0=(\chi^2_j)_{min}$ and $\nu=N_j$.

Two additional tests on each data set are possible. These tests are particularly useful in case that the overall reproduction of a data set is poor; they point to the culprit for the poor overall reproduction, namely to the shape or to 
the absolute normalisation of the data set.
\begin{itemize}
\item To examine the shape of the $j^{\rm th}$ data set (with respect to that of the BLS), it is important to allow the BLS to reproduce the data set `optimally', i.e., regardless of the scaling. This is equivalent to setting 
$\delta z_j \rightarrow \infty$ or $(\delta z_j)^{-2}=0$ in Eqs.~(\ref{eq:EQ07}) and (\ref{eq:EQ08}). The corresponding quantities will be denoted as $\hat{z}_j$ and $(\chi^2_j)_{stat}$, respectively; the quantity $(\chi^2_j)_{stat}$ 
represents the fluctuation in the $j^{\rm th}$ data set which (assuming the correctness of the shape of the data set) is of pure statistical nature.
\begin{equation} \label{eq:EQ09}
\hat{z}_j = \frac{\sum_{i=1}^{N_j} w_{ij} r_{ij}}{\sum_{i=1}^{N_j} w_{ij}}
\end{equation}
\begin{equation} \label{eq:EQ10}
(\chi^2_j)_{stat} = \frac{1}{\sum_{i=1}^{N_j} w_{ij}} \Big( \sum_{i=1}^{N_j} w_{ij} \sum_{i=1}^{N_j} w_{ij} r_{ij}^2 - \big( \sum_{i=1}^{N_j} w_{ij} r_{ij} \big)^2 \Big)
\end{equation}
As expected, both expressions are identical to those derived for the weighted average of a set of independent measurements, as well as for the corresponding $\chi^2$ value for constancy. Owing to the fact that the normalisation 
uncertainty is not used in Eq.~(\ref{eq:EQ10}), the quantity $(\chi^2_j)_{stat}$ is expected to follow the $\chi^2$ distribution with $\nu = N_j - 1$ DOF. The p-value, obtained from Eq.~(\ref{eq:EQ01}) with $x_0=(\chi^2_j)_{stat}$ and 
$\nu = N_j - 1$, may be used in order to assess the constancy of the input values $r_{ij}$ or, equivalently in our case, to examine the shape of the $j^{\rm th}$ data set with respect to the BLS~\footnote{In fact, the test simply assesses 
the goodness of the representation of the input data by one overall average value. A failure indicates either a bad shape (e.g., a slope being present in the input data) or `scattered' input values with small uncertainties. Visual 
inspection of the input data reveals that the latter option is not the case in the problem we have set out to investigate.}.
\item To assess the compatibility of the absolute normalisations of the $j^{\rm th}$ data set and of the BLS, one first estimates the scaling contribution to $(\chi^2_j)_{min}$ via the relation:
\begin{equation} \label{eq:EQ11}
(\chi^2_j)_{sc}=(\chi^2_j)_{min}-(\chi^2_j)_{stat}=\frac{(\delta z_j)^{-2} \big( \sum_{i=1}^{N_j} w_{ij} (r_{ij}-1) \big) ^2}{\big( \sum_{i=1}^{N_j} w_{ij} + (\delta z_j)^{-2} \big) \sum_{i=1}^{N_j} w_{ij}} \, \, \, ,
\end{equation}
where use of Eqs.~(\ref{eq:EQ08}) and (\ref{eq:EQ10}) has been made. The quantity $(\chi^2_j)_{sc}$ is expected to follow the $\chi^2$ distribution with $1$ DOF (which, of course, is the normal distribution).
\end{itemize}

We now summarise the three tests of the goodness of the reproduction of the $j^{\rm th}$ data set by a BLS.
\begin{itemize}
\item The overall reproduction can be tested using $(\chi^2_j)_{min}$ of Eq.~(\ref{eq:EQ08}) as $x_0$ in Eq.~(\ref{eq:EQ01}) and $\nu = N_j$ DOF.
\item The shape (statistical fluctuation) can be tested using $(\chi^2_j)_{stat}$ of Eq.~(\ref{eq:EQ10}) as $x_0$ in Eq.~(\ref{eq:EQ01}) and $\nu=N_j-1$ DOF.
\item The absolute normalisation can be tested using $(\chi^2_j)_{sc}$ of Eq.~(\ref{eq:EQ11}) as $x_0$ in Eq.~(\ref{eq:EQ01}) and $\nu=1$ DOF.
\end{itemize}

The tests outlined in the present section are objective. The only subjective aspect in the analysis pertains to the choice of the $\mathrm{p}_{min}$ level signifying statistical significance.

\section{\label{sec:Results}Results of the new analysis of the CHAOS DCSs}

We now present the results of the new analysis of the DENZ04 DCSs, using the changes in our approach as detailed in Sections \ref{sec:Introduction} and \ref{sec:Method}. We first investigated the description of the experimental data on 
the basis of the standard $K$-matrix low-energy parameterisations which we employ as first step in our PWAs, identifying and removing any outliers from the database; in this step of the optimisation, the theoretical constraint of 
crossing symmetry is not imposed onto the fitted scattering amplitudes. We used the split data sets, as defined in Refs.~\cite{mr1,abws}. For the truncated $\pi^+ p$ database, the minimal $\chi^2$ value was $397.4$ (to be compared to 
$401.2$ in Ref.~\cite{mr1}) for $260$ DOF. Exempting two data points which changed status~\footnote{The $19.90$ MeV, $42.75^\circ$ DCS, which was an outlier in Ref.~\cite{mr1}, became an accepted data point. On the contrary, the $32.00$ 
MeV, $40.83^\circ$ DCS, which was an accepted measurement earlier, turned into an outlier. Both points are close to the acceptance threshold and even small changes in the approach may affect their status.}, the list of outliers is the 
same as in Ref.~\cite{mr1}. The isospin-$\frac{3}{2}$ amplitudes were fixed from the final fit to the truncated $\pi^+ p$ database and were imported into the analysis of the $\pi^- p$ elastic-scattering database; only three data points 
were removed, the same measurements which had to be excluded in Ref.~\cite{mr1}. For the truncated $\pi^- p$ elastic-scattering database, the minimal $\chi^2$ value was $352.9$ (to be compared to $350.2$ in Ref.~\cite{mr1}) for $261$ 
DOF. Similarly to Ref.~\cite{mr1}, the result for $a^{cc}$, obtained in the final fit to the $\pi^- p$ elastic-scattering database, was unacceptable, about $0.042 \, m_c^{-1}$ ($m_c$ denotes the mass of the charged pion), differing 
from the experimental result obtained directly at the $\pi N$ threshold \cite{ss,orwmg} by a factor of $2$; this discrepancy is due to the inadequacy of the isospin-$\frac{3}{2}$ amplitudes to account simultaneously for both 
elastic-scattering reactions. The common fit to the truncated combined $\pi^\pm p$ elastic-scattering databases yielded no further outliers and a minimal $\chi^2$ value of $750.4$ (to be compared to $751.5$ in Ref.~\cite{mr1}) for $521$ 
DOF; the final $a^{cc}$ value was almost identical to the result we had obtained in the previous step, from the fit to the truncated $\pi^- p$ elastic-scattering database.

The fits of the ETH model to the truncated combined $\pi^\pm p$ elastic-scattering databases of the CHAOS Collaboration were next attempted, as described in Ref.~\cite{mr3}. The $\sigma$-meson mass $m_\sigma$ was varied within the 
interval which is currently recommended by the PDG \cite{pdg}. Unfortunately, none of these seven (see Ref.~\cite{mr3}, p. 178) fits terminated successfully. In all cases, negative diagonal elements were detected in the covariance matrix, 
urging the MINUIT software library \cite{jms} (FORTRAN version) to enforce positivity by adding arbitrary constants to the diagonal. As a result, the fitted values and the uncertainties of the model parameters are meaningless in all 
seven attempts to account for the DENZ04 DCSs on the basis of the ETH model.

We also followed the recommendation of the CHAOS Collaboration and analysed their unsplit (original) data sets. As explained in Ref.~\cite{mr1}, one expects that the problems, which we encountered in the analysis of their split data, 
can only be aggravated when using the unsplit data sets. To start with, $72$ data points were identified as outliers and had to be removed from the database: $65$ of these measurements belong to the $\pi^+ p$ data sets, $7$ to the 
$\pi^- p$ elastic-scattering ones. The result for $a^{cc}$, obtained in the final fit to the resulting truncated combined $\pi^\pm p$ elastic-scattering databases using our low-energy parameterisations of the $s$- and $p$-wave $K$-matrix 
elements, was equal to about $0.056 \, m_c^{-1}$. We subsequently attempted to fit the ETH model to the truncated combined $\pi^\pm p$ elastic-scattering databases. As in the case of the split data, all seven fits failed. Adding to the 
severity of the problems, as reported earlier when using the split data as input, the coupling constant $g_{\pi N N}$ came out in the vicinity of $0$ in all seven attempts. It is not possible to obtain anything reasonable from the unsplit 
data sets of the CHAOS Collaboration.

Given the seriousness of the problems we have encountered in analysing the DENZ04 DCSs, we can only investigate their reproduction on the basis of existing BLSs. The BLS in the present paper is obtained from the results of new fits to 
the truncated combined $\pi^\pm p$ elastic-scattering databases of Ref.~\cite{mr3}, after applying the changes as detailed in Section \ref{sec:Introduction}; the differences to the results of Ref.~\cite{mr3} were small, not exceeding $40 \%$ 
of the quoted uncertainties, save for the model parameter $\kappa_\rho$ where the difference amounts to about one standard deviation. The predictions for the `theoretical' values $y_{ij}^{th}$ and their uncertainties $\delta y_{ij}^{th}$ 
were obtained on the basis of $14$ million Monte-Carlo events for each data point in the CHAOS database. As always, the routine CORGEN of the CERN software library was used in the generation of the Monte-Carlo events; input to CORGEN is 
the `square root' of the covariance matrix, which was obtained from the optimisation results with the routine CORSET. The CPU consumption per energy and angle value was about $2$ min on a fairly-fast personal computer. As they are 
obtained in an analysis of a larger number of measurements, the uncertainties $\delta y_{ij}^{th}$ are generally expected to be smaller than the experimental uncertainties of single experiments. Nevertheless, they are sizeable in some 
kinematical regions covered by the CHAOS experiment, e.g., in backward $\pi^- p$ elastic scattering (see caption of Fig.~2 of Ref.~\cite{mr1}).

We now present the results of the analysis of the ratios $r_{ij}=y_{ij}^{exp}/y_{ij}^{th}$ for each data point in the DENZ04 database. Ideally, these ratios are constant (i.e., independent of $\theta$ and of $T$) and equal to $1$ 
(reflecting perfect agreement in the absolute normalisation of the data set being tested with respect to the BLS). Constant values of $r_{ij}$, different from $1$, point to differences in the absolute normalisation of the two sets of 
values, whereas a statistically-significant departure of $r_{ij}$ from constancy is evidence of a discrepancy in the shapes of the two angular distributions of the DCS. The ratios $r_{ij}$ for the DENZ04 $\pi^\pm p$ elastic-scattering 
databases are shown in Figs.~\ref{fig:PIPPE} and \ref{fig:PIMPE}.

The p-values of the reproduction of the DENZ04 DCSs \cite{chaos,denz} are given in Table \ref{tab:Reproduction1}; the table corresponds to the unsplit data sets. The eleven outliers, which had been identified in the first step of the 
analysis (i.e., employing standard low-energy parameterisations of the $s$- and $p$-wave $K$-matrix elements), have not been removed. The removal of these measurements induces very small effects and cannot alter the conclusions in the 
case of the $\pi^+ p$ data; as the free scaling (free floating) of the backward-angle $25.80$ MeV data set had been suggested in the first step of the optimisation, the treatment of the unsplit $\pi^+ p$ $25.80$ MeV data set is not 
straightforward. The removal of the two outliers from the $\pi^- p$ $25.80$ MeV data set improves its reproduction by the BLS; the problem with this data set mainly rests with the peculiar shape of the forward-angle measurements (the 
same behaviour is observed in the corresponding $\pi^+ p$ data set), where the dominant contribution to the $\pi N$ scattering amplitude comes from the electromagnetic interaction.

All six p-values for the overall reproduction of the $\pi^+ p$ data sets in Table \ref{tab:Reproduction1} are smaller than the significance level $\mathrm{p}_{min}$ for the rejection of the null hypothesis which we adopt in our PWAs, 
namely $\mathrm{p}_{min} \approx 1.24 \cdot 10^{-2}$; this value of $\mathrm{p}_{min}$, equivalent to a $2.5\sigma$ effect in the normal distribution, is close to $1 \%$ which most statisticians adopt in the statistical hypothesis 
testing. The $37.10$ MeV $\pi^+ p$ data set is the best reproduced by the BLS ($\mathrm{p} \approx 3.65 \cdot 10^{-3}$), whereas the $25.80$ MeV $\pi^+ p$ data set is obviously the worst reproduced. Beyond doubt, the problems in the 
reproduction of these data sets relate to their shape. On the other hand, the only $\pi^- p$ $r_{ij}$ ratios, which are poorly reproduced, are those of the $25.80$ MeV data set, which appears to contain suspicious measurements at small 
$\theta$; the reproduction of the remaining $\pi^- p$ elastic-scattering data sets is satisfactory. Evidently, the angular distribution of the DENZ04 $\pi^+ p$ database disagrees (in shape) with the rest of the $\pi^+ p$ database, 
whereas the DENZ04 $\pi^- p$ elastic-scattering DCSs appear to be in reasonable agreement with the BLS. We had reached the same conclusion in Ref.~\cite{mr1}. Regarding the split data, four (out of $17$) $\pi^+ p$ data sets are poorly 
reproduced, as is the forward-angle $25.80$ MeV $\pi^- p$ elastic-scattering data set. Given that our results now contain all contributions to the uncertainties, there is hardly room for improvement in the reproduction of the 
DENZ04 $\pi^+ p$ database.

\section{\label{sec:Conclusions}Discussion and conclusions}

In the present paper, we analysed the differential cross sections (DCSs) of the CHAOS Collaboration \cite{chaos,denz} and investigated their reproduction on the basis of the results obtained from a partial-wave analysis of the rest of 
the low-energy (pion laboratory kinetic energy $T \leq 100$ MeV) $\pi^\pm p$ elastic-scattering data \cite{mr3}. Since our previous analysis of the same data \cite{mr1} appeared, there have been three main developments calling for a new 
investigation of this subject.
\begin{itemize}
\item When investigating the reproduction of an experiment on the basis of a given `theoretical' solution, our results also include now the uncertainties $\delta y_{ij}^{th}$ of the theoretical values (see Eq.~(\ref{eq:EQ04})).
\item Recent developments regarding the proton electromagnetic form factors suggest the replacement of the forms we used in our earlier analyses; from now on, we will adopt the parameterisation (and the optimal parameter 
values) of Ref.~\cite{vamx}. Additionally, the pion electromagnetic form factor will be parameterised via a monopole form.
\item The fits of the ETH model \cite{glmbg,mr3} to the experimental data now involve the variation of the $\sigma$-meson mass ($m_\sigma$) within the interval which is recommended by the Particle-Data Group \cite{pdg} (currently, 
between $400$ and $550$ MeV).
\end{itemize}
The new analysis of the CHAOS DCSs demonstrated that the conclusions of Ref.~\cite{mr1} hold and that the angular distribution of their $\pi^+ p$ cross sections is in conflict with the rest of the modern (meson-factory) low-energy 
$\pi^+ p$ database.

One might argue that a disagreement between any two sets of data attests only to the faultiness of at least one of them. In order to determine which of the two sets of values is flawed, the insight gained from theory is often valuable. 
To this end, we analysed the two sets of measurements, irrespective of one another, by employing two theoretical approaches in the analysis, namely the standard low-energy parameterisations of the $s$- and $p$-wave $K$-matrix elements 
and the ETH model. We are critical of the DCSs of the CHAOS Collaboration because we have not been able to obtain anything reasonable from the analysis of these data following \emph{either} theoretical approach. On the contrary, we 
did not encounter problems when analysing the rest of the modern database following \emph{both} theoretical approaches. As the beam energy in the CHAOS experiments was sufficiently low, an investigation of the description of their DCSs 
within the framework of the Chiral Perturbation Theory (e.g., with the method of Ref.~\cite{aco}) should be possible. This would be an interesting subject to pursue.

%

\newpage
\begin{table}
{\bf \caption{\label{tab:Reproduction1}}}The details of the reproduction of the differential cross sections of the CHAOS Collaboration \cite{chaos,denz} by the baseline solution (BLS) obtained following the approach of Ref.~\cite{mr3}. 
The results now include the uncertainties $\delta y_{ij}^{th}$ of the BLS (see Eq.~(\ref{eq:EQ04})). The columns represent: the pion laboratory kinetic energy $T$ (in MeV), the number of data points $N_j$ of the $j^{\rm th}$ experimental 
data set, and the three p-values corresponding a) to the overall reproduction of the data set, b) to the reproduction of its shape, and c) to the reproduction of its absolute normalisation. The table corresponds to the unsplit (original) 
data sets of the CHAOS Collaboration.
\vspace{0.2cm}
\begin{center}
\begin{tabular}{|c|c|c|c|c|}
\hline
$T$ & $N_j$ & Overall & Shape & Absolute normalisation \\
\hline
\multicolumn{5}{|c|}{$\pi^+ p$ scattering} \\
\hline
$19.90$ & $33$ & $2.37 \cdot 10^{-9}$ & $1.30 \cdot 10^{-9}$ & $8.46 \cdot 10^{-1}$ \\
$25.80$ & $43$ & $9.97 \cdot 10^{-78}$ & $3.28 \cdot 10^{-78}$ & $6.00 \cdot 10^{-1}$ \\
$32.00$ & $46$ & $2.18 \cdot 10^{-4}$ & $1.53 \cdot 10^{-4}$ & $9.67 \cdot 10^{-1}$ \\
$37.10$ & $49$ & $3.65 \cdot 10^{-3}$ & $4.99 \cdot 10^{-3}$ & $1.00 \cdot 10^{-1}$ \\
$43.30$ & $53$ & $3.19 \cdot 10^{-11}$ & $1.87 \cdot 10^{-11}$ & $8.65 \cdot 10^{-1}$ \\
$43.30$(rot.) & $51$ & $1.16 \cdot 10^{-12}$ & $7.22 \cdot 10^{-13}$ & $5.81 \cdot 10^{-1}$ \\
\hline
\multicolumn{5}{|c|}{$\pi^- p$ elastic scattering} \\
\hline
$19.90$ & $31$ & $9.90 \cdot 10^{-1}$ & $9.89 \cdot 10^{-1}$ & $4.38 \cdot 10^{-1}$ \\
$25.80$ & $45$ & $4.82 \cdot 10^{-11}$ & $3.76 \cdot 10^{-11}$ & $3.30 \cdot 10^{-1}$ \\
$32.00$ & $45$ & $1.40 \cdot 10^{-2}$ & $5.26 \cdot 10^{-2}$ & $4.31 \cdot 10^{-3}$ \\
$37.10$ & $50$ & $3.55 \cdot 10^{-1}$ & $4.08 \cdot 10^{-1}$ & $1.17 \cdot 10^{-1}$ \\
$43.30$ & $51$ & $1.44 \cdot 10^{-1}$ & $2.47 \cdot 10^{-1}$ & $2.12 \cdot 10^{-2}$ \\
$43.30$(rot.) & $49$ & $6.38 \cdot 10^{-1}$ & $8.71 \cdot 10^{-1}$ & $5.27 \cdot 10^{-3}$ \\
\hline
\end{tabular}
\end{center}
\end{table}

\clearpage
\begin{figure}
\begin{center}
\includegraphics [width=15.5cm] {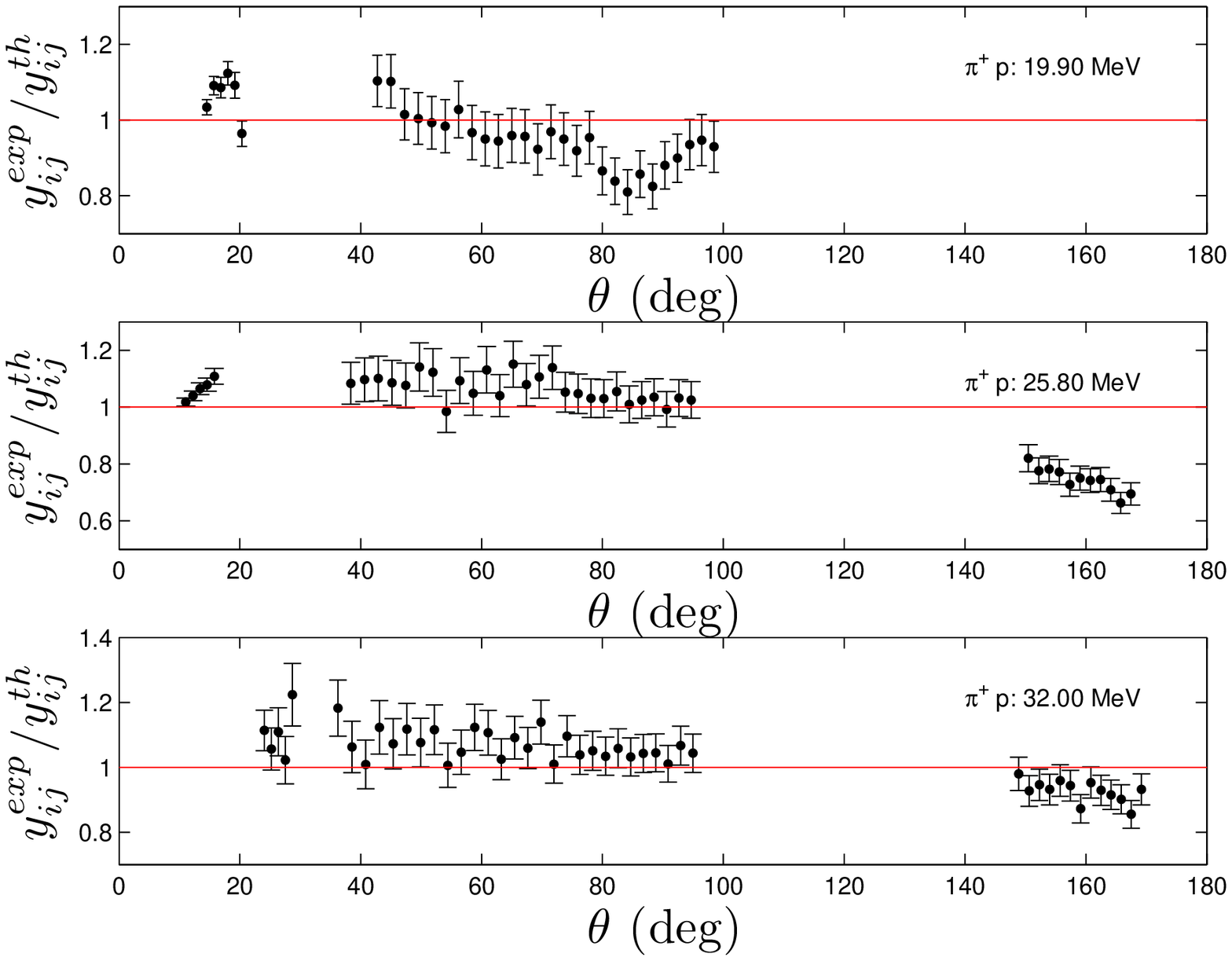}
\end{center}
\end{figure}

\clearpage
\begin{figure}
\begin{center}
\includegraphics [width=15.5cm] {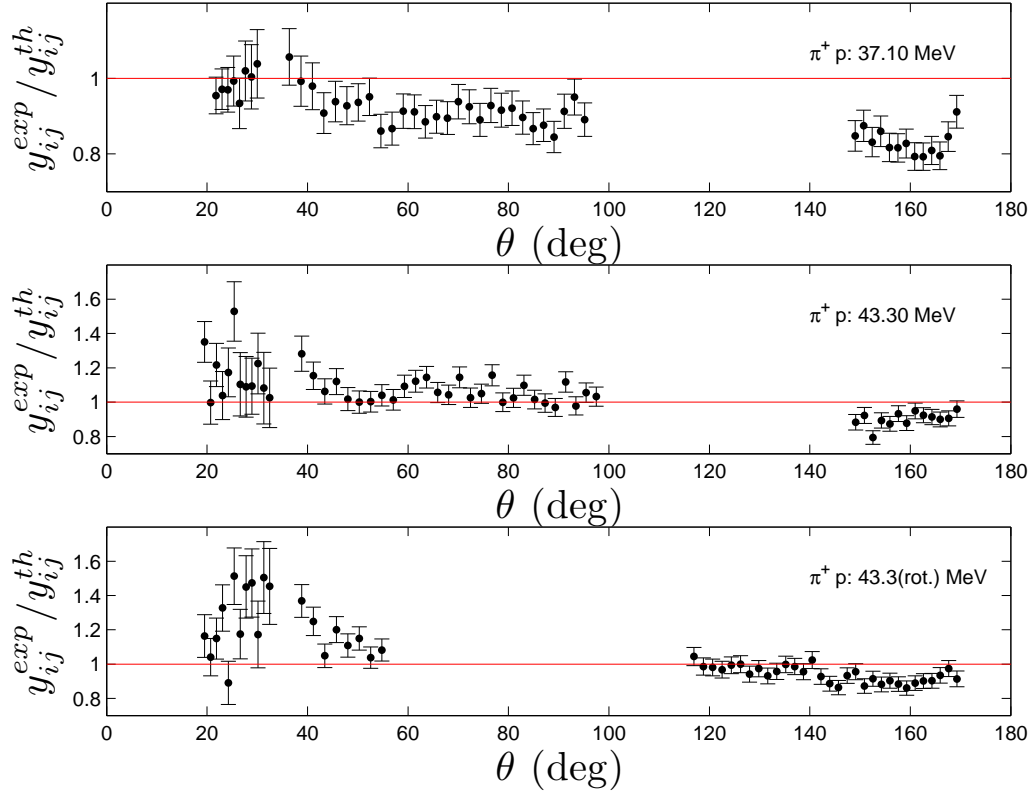}
\caption{\label{fig:PIPPE}The $\pi^+ p$ differential cross sections of the CHAOS Collaboration \cite{chaos,denz} ($y_{ij}^{exp}$), normalised to the corresponding predictions ($y_{ij}^{th}$) obtained following the approach of 
Ref.~\cite{mr3}; the eight outliers in the $\pi^+ p$ database are also contained in the figure. The normalisation uncertainties of the experimental data sets (see Refs.~\cite{chaos,mr1} for details) are not shown. Unlike Ref.~\cite{mr1} 
(e.g., see caption of Fig.~1 therein), the uncertainties $\delta y_{ij}^{th}$ are now included in the results (see Eq.~(\ref{eq:EQ04})).}
\end{center}
\end{figure}

\clearpage
\begin{figure}
\begin{center}
\includegraphics [width=15.5cm] {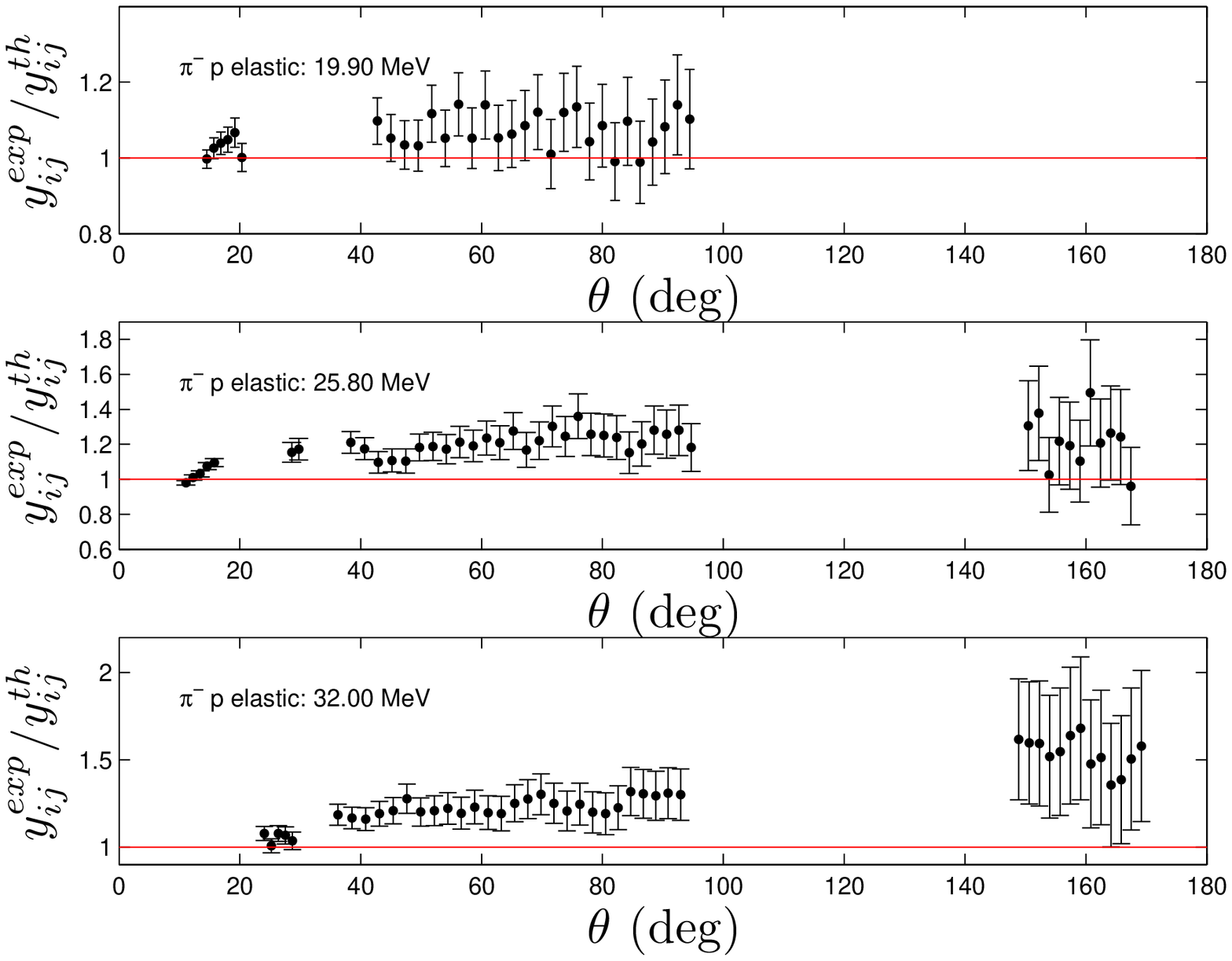}
\end{center}
\end{figure}

\clearpage
\begin{figure}
\begin{center}
\includegraphics [width=15.5cm] {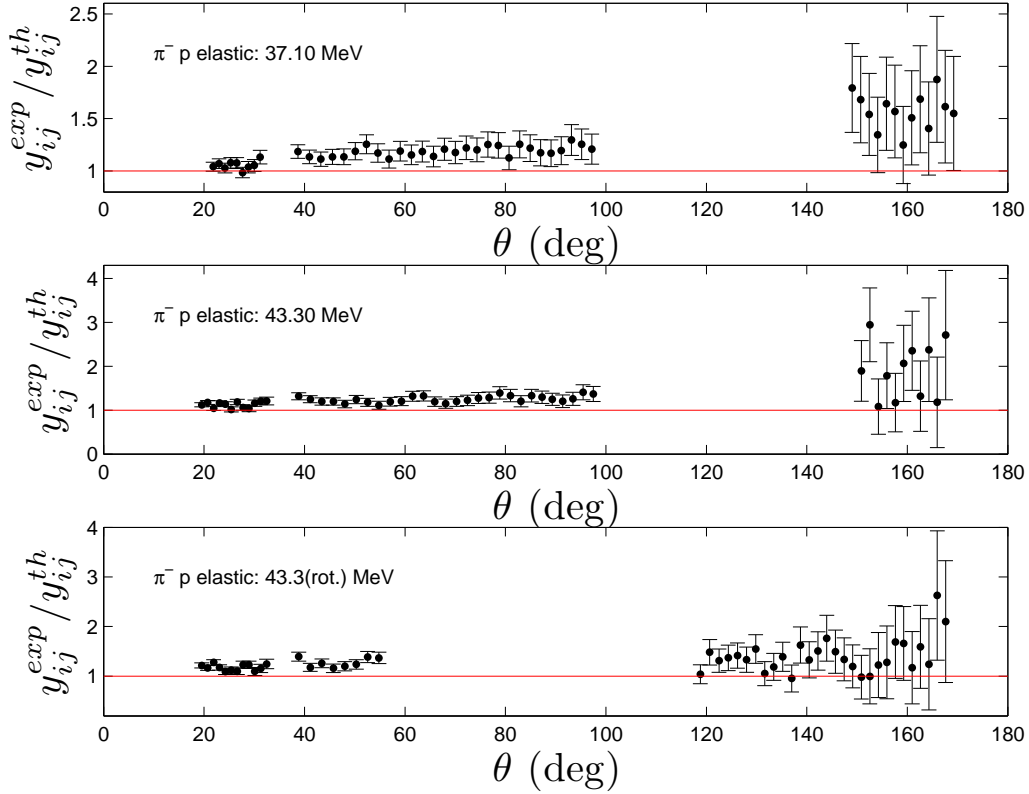}
\caption{\label{fig:PIMPE}The $\pi^- p$ elastic-scattering differential cross sections of the CHAOS Collaboration \cite{chaos,denz} ($y_{ij}^{exp}$), normalised to the corresponding predictions ($y_{ij}^{th}$) obtained following the 
approach of Ref.~\cite{mr3}; the three outliers in the $\pi^- p$ elastic-scattering database are also contained in the figure. The normalisation uncertainties of the experimental data sets (see Refs.~\cite{chaos,mr1} for details) are 
not shown. Unlike Ref.~\cite{mr1} (see caption of Fig.~2 therein), the uncertainties $\delta y_{ij}^{th}$, which are sizeable at backward angles, are now included in the results (see Eq.~(\ref{eq:EQ04})).}
\end{center}
\end{figure}

\end{document}